\documentclass[10pt,letterpaper]{article}
\usepackage[top=1in,left=1in,right=1in,bottom=1in]{geometry}
\usepackage{amsmath,amssymb}
\usepackage{changepage}
\usepackage{textcomp,marvosym}
\usepackage{cite}
\usepackage{nameref,hyperref}
\usepackage{microtype}
\DisableLigatures[f]{encoding = *, family = * }
\usepackage[table]{xcolor}
\usepackage{array}
\usepackage{subcaption}
\usepackage{tikz}
\usepackage{algorithm}
\usepackage{algpseudocode}
\usepackage{float}

\usepackage{xcolor}
\usepackage{setspace}
\newcommand{\blue}[1]{#1}

\newcolumntype{+}{!{\vrule width 2pt}}
\newlength\savedwidth

\raggedright
\usepackage[aboveskip=1pt,labelfont=bf,labelsep=period,justification=raggedright,singlelinecheck=off]{caption}

\makeatletter
\renewcommand{\@biblabel}[1]{\quad#1.}
\makeatother

\usepackage{lastpage,fancyhdr,graphicx}
\usepackage{epstopdf}
\begin{document}
%\doublespacing  % removed for arXiv preprint
\vspace*{0.2in}

\begin{flushleft}
{\Large
\textbf\newline{Opinion polarization from compression-based decision making where agents optimize local complexity and global simplicity}
}
\newline
Alina Dubovskaya\textsuperscript{1,2,3*},
David J. P. O'Sullivan\textsuperscript{2},
Michael Quayle\textsuperscript{1,3}
\\
\bigskip
\textbf{1} Centre for Social Issues Research, Department of Psychology, University of Limerick, Limerick, V94 T9PX, Ireland\\
\textbf{2} Mathematics Applications Consortium for Science and Industry (MACSI), Department of Mathematics and Statistics, University of Limerick, Limerick, V94 T9PX, Ireland\\
\textbf{3} Lero, the Research Ireland Centre for Software, University of Limerick, Limerick, V94 T9PX, Ireland\\
\bigskip
* alina.dubovskaya@ul.ie
\end{flushleft}

\section*{Abstract}
Understanding social polarization requires integrating insights from psychology, sociology, and complex systems science. Agent-based modeling provides a natural framework to combine perspectives from different fields and explore how individual cognition shapes collective outcomes. This study introduces a novel agent-based model that integrates two cognitive and social mechanisms: the desire to be unique within a group (optimal distinctiveness theory) and the tendency to simplify complex information (cognitive compression). \blue{In the model, virtual agents interact in pairs and decide whether to adopt each other's opinions by balancing two opposing drives: maximizing opinion diversity within their local social group while simplifying the overall opinion landscape, with both evaluated using Shannon entropy.} We show that the combination of these mechanisms can reproduce real-world patterns, such as the emergence of distinct heterogeneous opinion clusters. Moreover, unlike many existing models where opinions become fixed once opinion groups form, individuals in our model continue to adjust their opinions after clusters emerge, leading to ongoing variation within and between opinion groups. Computational experiments reveal that \blue{polarization emerges when local group sizes are moderate (consistent with Dunbar’s number)}, while smaller groups cause fragmentation and larger ones hinder distinct cluster formation. Higher cognitive compression increases unpredictability, while lower compression produces more consistent group structures. These results demonstrate how simple psychological rules can generate complex, realistic social behavior and advance understanding of polarization in human societies.

%\linenumbers

\section*{Introduction}
In today’s interconnected world, people's opinions play a powerful role in shaping the social, political, and economic landscape. How these opinions are distributed across a population can influence public policy, cultural norms, and the success or failure of businesses. Political and ideological polarization is seemingly becoming increasingly pronounced, raising concerns about its impact on democratic institutions and social cohesion \cite{dellaposta2020pluralistic,Graham2020,Arbatli2021,Zhou2016}.
\blue{Many social mechanisms have been identified as contributing to opinion polarization, including elite cueing \cite{druckman2013elite}, socio-political identity sorting \cite{mason2018ideologues}, shifts in media environments \cite{kubin2021role}, and socio-economic inequality \cite{mutz2018status}, to name a few. However, how polarization emerges as a dynamic, self-organizing process from individual-level interactions in complex social systems remains less understood.}  Insights from social psychology, political science, and complexity science may be necessary to uncover the mechanisms through which such emergent polarization arises.

Societies and social groups are complex systems where various mechanisms operate on different scales, ranging from neurological mechanisms (such as cognitive limitations and biases) and individual reasoning and preferences to external factors (such as the information environment in which an individual operates)~\cite{davis2015theories,gavrilets2024modelling,galesic2023beyond}. Mechanisms that work at the level of local interaction are among the major factors shaping the distribution of opinions in a population~\cite{centola2015spontaneous,castellano2009statistical,warncke2025country}.

Various approaches are employed to study the complex system of opinion formation, including behavioral experiments~\cite{luders2024attitude,o2024strategic,miranda2024indirect}, survey research~\cite{maher2020mapping,carpentras2022mapping,dinkelberg2021multidimensional,chen2025broken}, data analysis of online social platforms~\cite{karsai2016local, pena2025finding}, as well as mathematical and computational modeling~\cite{deffuant2000mixing, degroot1974reaching, axelrod1997dissemination,galesic2021integrating,friedkin2016network,parsegov2015new}. Among computational approaches, agent-based models (ABMs) have become a popular tool for testing basic mechanisms that may underlie complex systems.
In these models, we prescribe mechanisms that may facilitate decision-making at the individual level during interactions and aim to study how these micro-level mechanisms can lead to the formation of emergent population-level phenomena such as consensus or polarization of opinions~\cite{flache2017models,vallacher2017computational}.

Several well-established ABMs have demonstrated that polarization can emerge from individual-level mechanisms such as bounded confidence in the  Deffuant-Weisbuch~\cite{deffuant2000mixing} and the Hegselmann-Krause~\cite{hegselmann2005opinion} models, cultural similarity in the Axelrod's model of Cultural Dissemination~\cite{axelrod1997dissemination} and by averaging neighboring opinions in the DeGroot model~\cite{degroot1974reaching}. In these models, opinion group formation arises as a result of limited interaction across different opinion groups and they typically converge to simplistic opinion distributions—either global consensus or a small number of homogeneous clusters where individuals hold same or near the same opinions. Once the opinion clusters are formed, agents stop updating their opinions entirely, becoming locked in rigid opinion groups.  In contrast, real-world opinion landscapes are rarely so discrete. Even in strongly polarized societies, individuals within the same group do not hold identical beliefs; instead, there is always a degree of variation and ongoing opinion shifts. Introducing noise into these models addresses some of these limitations and leads to more complex behavior~\cite{pineda2009noisy,pineda2011diffusing}. Nevertheless, in their pure form, the mechanisms incorporated in these models lead to oversimplified, static outcomes.

Recent works in social psychology look at polarization as a fundamental process by which groups manage their identity~\cite{bliuc2021online, Quayle10062025, Durrheim06052025,smith2024polarization}. Polarized social systems allow individuals to easily differentiate between in-group and out-group members, recognize their own belonging, and signal their identity. However, even while being motivated to affiliate with clearly differentiable groups, people still value individuality within those groups ~\cite{snyder2012uniqueness,brewer1991social}. 
Theories of uniqueness~\cite{snyder2012uniqueness} and individuation~\cite{codol1975so,lemaine1974social} suggest that individuals strive to balance two opposing drives: the desire to belong to a well-defined social group and the desire to feel unique~\cite{pickett2006using,postmes2006individuality}. Thus, individuals prefer attitudinal positions that are close enough to their group to signal affiliation, yet distinct enough to assert personal identity~\cite{rubin2012}. In other words, when choosing an attitude or opinion, people favor positions that are locally complex but globally comprehensible. Moreover, optimal distinctiveness theory~\cite{brewer1991social,leonardelli2010optimal} argues that these two competing desires lead to social groups that are simultaneously sufficiently inclusive and sufficiently distinct, satisfying both needs in an intricate balance. Certainly, real-world polarization is characterized not by perfect homogeneity within groups, but by stable clusters that retain internal variability and allow for fluidity in opinion within group limits. In this paper, we propose that the desire for optimal distinctiveness may serve as a fundamental mechanisms for polarization and also explain the persistence of opinion diversity within cohesive social groups.

A second mechanism we investigate is cognitive compression. Originating from early work by Lippmann~\cite{lippmann1965public}, this concept highlights how humans simplify complex information into simplified cognitive representations—such as stereotypes or social categories, to process new information more efficiently. Experimental studies using the ``telephone game'' (i.e., transmission-chain experiments) have shown that people perceive and reproduce information in increasingly compressed forms as it passes through communication chains~\cite{tamariz2015culture}. These findings support the observation that humans can comprehend and act in a complex world because they encode it in compressed forms that make it easy to categorize others into groups and identify social outliers~\cite{chaitin2006limits,Durrheim06052025}.

In this paper, we develop a novel agent-based model in which individuals seek to balance their need for uniqueness within their social group with their tendency to simplify the broader opinion environment. These two competing desires work in opposite directions since greater local complexity limits the degree to which the opinion space can be simplified, and thus we expect the model to produce complex behavior. \blue{The central question of this study is: what population-level opinion patterns emerge when agents simultaneously optimize for local complexity and global simplicity, and under what conditions do these mechanisms lead to polarization?}
We show that these mechanisms can produce:
a) population-level opinion group formation,
b) sustained variability of opinions within social groups, and
c) individual-level mobility within social groups, i.e., shifting positions within or between groups.
By modeling these dynamics, we offer a new theoretical lens on polarization that explains not only the formation of distinct opinion groups, but also why diversity and movement persist within those groups in social systems.

\section*{Model of balancing local complexity and global simplicity in the opinion distribution}
In the proposed model, agents interact in pairs and have an opportunity to swap their opinions to the opinion of the second interacting agent during Monte Carlo simulations. To decide whether to adopt the second agent's opinion or to keep their current one, the agent evaluates whether the new opinion would place them in a better position within their local social group \blue{or make the overall opinion distribution in the population more comprehensible (or both)}. A better position within the local group means increased diversity within that group; a more comprehensible global distribution leads to easily identifiable clusters in the population.

\blue{Formally, we have $N$ agents in the model that hold a univariate opinion ranging between
0 and 1. The opinion profile at time $t$ is denoted by $X(t)=\{x_i(t)\}_{i=1}^{N}$, where
$t$ is the time step of the simulation. Initially, the agents are assigned random opinions drawn from the continuous uniform distribution, $X(0)\sim \text{Uniform}(0, 1)$. For each agent $i$ we also define a subset, $L_i(t)\subseteq X(t)$, which contains the opinions of the agents that agent $i$ considers their local social group. How opinions are chosen for the local group will be explained below.}

\begin{figure}[H]
\centering
\includegraphics[width=0.75\textwidth]{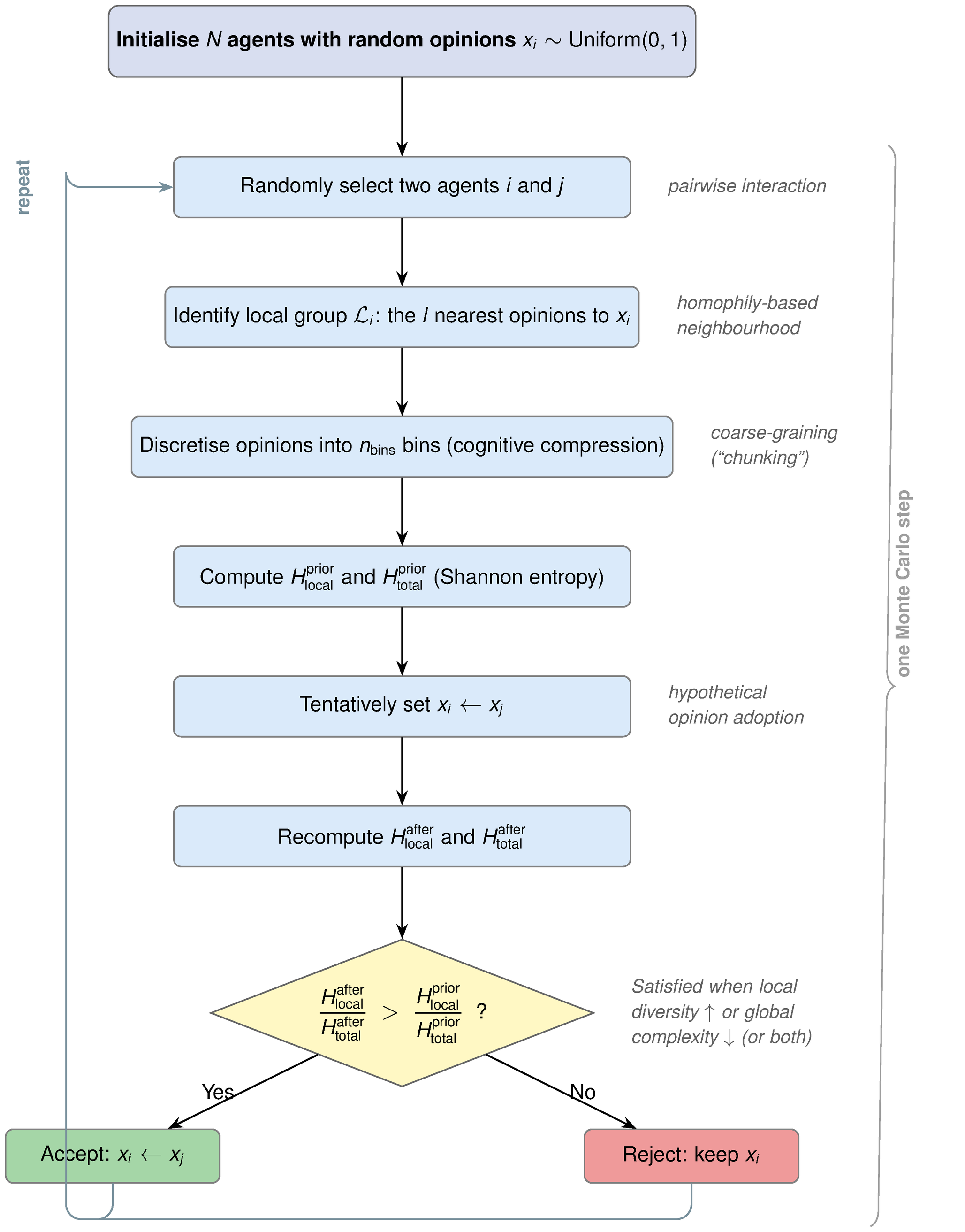}
\caption{\blue{Flow chart of the opinion update rule. At each Monte Carlo step, two agents are randomly selected, and the first agent tentatively adopts the second agent's opinion. The change is accepted if it increases the ratio of local to global Shannon entropy, meaning either local opinion diversity increases or global complexity decreases (or both).}}
\label{fig:flowchart}
\end{figure}

\blue{We then perform Monte Carlo simulations where, at each round, two agents are randomly selected for interaction, and the first agent has an opportunity to change their opinion to that of the second agent. To decide whether adopting a new opinion will position them ``better'' in the opinion space, the agent calculates the complexity of information in the opinion distribution within their local group and the entire population, both before the interaction and after the hypothetical change, using the Shannon entropy~\cite{shannon1948mathematical}}
\blue{Specifically, at each time step of the Monte Carlo simulations:
\begin{enumerate}
    \item two agents are selected at random;
    \item the first agent considers the second agent's opinion; and if accepting the new opinion would increase the first agent's ratio of local to global entropy relative to their old one, they do so. Otherwise, they keep their current opinion.
\end{enumerate}
The simulation continues until the bailout time, which allows sufficient time for the long-term behavior to stabilize.}
\blue{Fig~\ref{fig:flowchart} illustrates the main steps of the opinion update procedure within a single Monte Carlo step.}

\blue{We now explain in detail how we calculate the entropies. When agent $i$ interacts with agent $j$, we calculate the Shannon entropy across the sets $X(t)$ and $L_i(t)$ and refer to them as the total and local entropy prior to interaction. We then replace the opinion of agent $i$ with the opinion of agent $j$ and calculate the total and local entropy in the case that agent $i$ decides to switch opinion, and compare their ratio. The opinion update rule then is:
\begin{equation}
    x_i(t+1)=\begin{cases}
                x_{j}(t) & \quad \text{if}\quad \dfrac{\mathrm{H_{\text{local}}^{after}}}{\mathrm{H_{\text{total}}^{after}}}>
                \dfrac{\mathrm{H_{\text{local}}^{prior}}}{\mathrm{H_{\text{total}}^{prior}}}, \\
                x_{i}(t) & \quad \text{otherwise.}
            \end{cases}
            \label{eq:opinion_update}
\end{equation}
where $H_{\text{local}}^{\text{prior}}$ and $H_{\text{total}}^{\text{prior}}$ represent the entropy of opinions within the agent's local group and the entire population, respectively, prior to the interaction, while $H_{\text{local}}^{\text{after}}$ and $H_{\text{total}}^{\text{after}}$ are the corresponding entropies in the case that agent $i$ decides to change their opinion.}
Inequality~\eqref{eq:opinion_update} holds when either the local entropy increases meaning more variability within the local group or the global entropy decreases that indicates a simplified global opinion distribution. \blue{Algorithm~\ref{alg:opinion_update} shows the pseudocode for the opinion updating rule.}

\blue{To calculate Shannon entropy from the continuous distributions, we first divide the opinion space into $n_{\text{bins}}$ discrete intervals, or bins, and calculate the probability distribution across the binned distribution, $p_k$. The entropy is then evaluated as
\begin{equation}
    \mathrm{H(X)} =-\sum_{k=1}^{n_\text{bins}} p_k\log_{2}p_k, \quad \text{where}\quad p_k=\frac{n_k}{N},
\end{equation}
where $n_k$ is the number of opinions in the $k$-th bin. Algorithm~\ref{alg:entropy} illustrates the algorithm for entropy calculation.}
\blue{This binning process reflects cognitive compression that agents apply when perceiving the opinion distribution in the population. Similar opinions are grouped into the same bin, rendering them indistinguishable for the purpose of entropy evaluation. The number of bins, $n_{\text{bins}}$, reflects the level of cognitive compression: a larger $n_{\text{bins}}$ allows more detailed processing, while a smaller $n_{\text{bins}}$ leads to more compressed, generalized information.}

\blue{These bins operationalize psychological "chunking", a core cognitive mechanism for organizing information across tasks and domains~\cite{gobet2001chunking}. Classic work by Miller~\cite{miller1956magical} demonstrated that people can reliably maintain roughly $7\pm2$ chunks of information. Subsequent studies have shown this to be on the higher side of human performance in most tasks; but that the optimal number of chunks remains relatively constant regardless of the complexity or level of abstraction of the task and that memory performance benefits from information compression similarly across both simple and complex material~\cite{mathy2018simple}. Thus, we assume that agents use the same level of information compression regardless of the level of abstraction being applied, whether chunking fine-grained (opinions of the local group) or coarse-grained (opinions of the whole population) information.
Translating this to our simulation, each agent encodes information, whether about the opinions in their local social group or in the entire population, into a roughly constant number of cognitive units, represented by the number of bins, which naturally yields finer-grained representations for smaller, more homogeneous opinions of the local groups and coarser-grained representations for opinions of the larger, more heterogeneous population.}

A compression mechanism based on coarse- and fine-grained representations can be compared with the example of observing a crowd versus an individual: when observing an individual, we perceive finer details; when viewing a crowd, we see only general patterns.
The \textbf{number of bins used in entropy calculation}, $n_{\text{bins}}$, is our first control parameter.

\begin{algorithm}[t]
\caption{Opinion update rule}
\label{alg:opinion_update}
\begin{algorithmic}[1]

\State Initialize opinions $\{x_k\}$ for all agents $k$. Define the set of all opinions as $\mathcal{X}$.

\While{stopping criterion not met}

    \State Select two agents $i$ and $j$ uniformly at random

    \State Construct local group $\mathcal{L}_i$ consisting of $l$ nearest opinions of opinion $x_i$

    \State Discretize opinions in $\mathcal{L}_i$ into $n_{\text{bins}}$ bins to obtain opinion counts set~$\{n_b\}_{\mathcal{L}_i}, \; b=1,\dots n_{\text{bins}}$, where $n_b$ is the number of opinions in bin $b$

    \State Compute
    $H_{\text{local}}^{\text{prior}}$ from $\{n_b\}_{\mathcal{L}_i}$ with algorithm~\ref{alg:entropy} (below)

    \State Discretize opinions in the whole opinion space $\mathcal{X}$ into $n_{\text{bins}}$ bins to obtain opinion counts set $\{n_b\}_{\mathcal{X}}, \; b=1,\dots n_{\text{bins}}$

    \State Compute
    $H_{\text{total}}^{\text{prior}}$ from $\{n_b\}_{\mathcal{X}}$ with algorithm 2

    \State Temporarily substitute $x_i \leftarrow x_j$

    \State Repeat steps 4 and 5 for the updated opinions set to construct opinion count set $\{n_b\}'_{\mathcal{L}_i}$ accounting for temporarily changed opinion $x_i$

    \State Repeat steps 7 and 8 for the updated opinions set to construct opinion count set $\{n_b\}'_{\mathcal{X}}$ accounting for temporarily changed opinion $x_i$

    \State Compute
    $H_{\text{local}}^{\text{after}}$ from $\{n_b\}'_{\mathcal{L}_i}$ and $H_{\text{total}}^{\text{after}}$ from $\{n_b\}'_{\mathcal{X}}$ with algorithm 2

    \If{
    $\dfrac{H_{\text{local}}^{\text{after}}}{H_{\text{total}}^{\text{after}}}
    >
    \dfrac{H_{\text{local}}^{\text{prior}}}{H_{\text{total}}^{\text{prior}}}$
    }

        \State Accept update: $x_i \leftarrow x_j$

    \Else

        \State Reject update: keep $x_i$ unchanged

    \EndIf

\EndWhile

\end{algorithmic}
\end{algorithm}

\begin{algorithm}[t]
\caption{Shannon entropy of discretised opinions}
\label{alg:entropy}
\begin{algorithmic}[1]

\Require Discretised opinion counts $\{n_b\}$
\Require Total number of opinions $N = \sum_{b=1}^{n_{\text{bins}}} n_b$

\Ensure Shannon entropy $H$

\State $H \leftarrow 0$

\For{$b = 1$ to $n_{\text{bins}}$}

    \If{$n_b > 0$}

        \State $p_b \leftarrow \dfrac{n_b}{N}$

        \State $H \leftarrow H - p_b \log(p_b)$

    \EndIf
\EndFor

\State \Return $H$

\end{algorithmic}
\end{algorithm}

We now explain how we determine which agents belong to a given agent’s local group, i.e. how $L_i$ set is defined. Several definitions could be motivated by social psychology; however, for simplicity, we define the local group as the set of individuals whose opinions are closest to the current agent’s opinion. This choice is motivated by the concept of \textit{homophily}—the tendency of individuals to associate with like-minded others.

The \textbf{size of the local group}, denoted by $l$, becomes our second control parameter. A larger $l$ represents broader social groups to which individuals typically belong, while a smaller $l$ indicates smaller social groups. By varying $l$, we can explore how the size of agents' social groups affects the population-level distribution of opinions. \blue{A complete description of the model following the ODD (Overview, Design concepts, Details) protocol is provided in \nameref{S1_Appendix}.}

\subsection*{Parameter selection}

\blue{To choose plausible values for the parameters representing the model’s demographic structure, i.e., the size of the local social groups and the total population size, we look to empirical findings on the typical scale of human social organization.
Ethnographic primate research associated with Dunbar’s social brain hypothesis~\cite{dunbar2010bondedness,dunbar1998social} suggests that the size of social groups in which people can maintain stable social knowledge of one another, supporting the formation and reinforcement of shared norms through repeated interaction, is approximately 100--200 individuals~\cite{dunbar1992neocortex,shultz2010encephalization,barton1996neocortex,lewis2011ventromedial,powell2012orbital,kanai2012online}. Empirical studies of offline and online social networks suggest that this number typically lies between 100 and 300~\cite{gonccalves2011modeling,hernando2010unravelling}. This scale is therefore a suitable range for the individual's immediate social environment, i.e., what in the model is called a local social group of an agent.
}

\blue{At the same time, archaeological studies of early human populations show that once settlements aggregate beyond this threshold, they expand into populations of several hundreds to low thousands, accompanied by the appearance of institutional or organizational mechanisms that stabilize cooperation among partially unfamiliar individuals \cite{Johnson1982,Kuijt2000}. Such supra-Dunbar populations correspond to what can be describe as tribal or regional network scales ($\approx500-2500$ people), where social cohesion is no longer maintained solely by direct personal ties but by mesoscale structures (e.g., subgroups, norms, and signaling systems) \cite{Johnson1982}.}

\blue{The total number of agents in our study is fixed at $N=1000$, which is consistent with these estimates of regional population size. This number is also a common choice in agent-based models of opinion dynamics. Model sizes of this order are typically sufficient to avoid finite-size effects while remaining computationally efficient to allow the investigation of emergent collective patterns across repeated simulations~\cite{castellano2009statistical,bonabeau2002agent,deffuant2000mixing}. Strictly speaking, the population size is an additional control parameter; however, due to the computational cost of the simulations, it was not explored systematically. It is worth noting that, instead of considering absolute values, one can use ratios of the local group size to the total population size, as well as the number of bins relative to the population size. This approach ensures that the control parameters remain independent of changes in population size. In that case, the absolute value of the population size would not matter, as long as it is sufficiently large. However, for the sake of clearer presentation of the results, we retain absolute values of the control parameters.}

To motivate a plausible range of values for the number of bins used in entropy calculations, $n_{\text{bins}}$, we draw on results from cognitive psychology on information compression and short-term memory capacity~\cite{miller1956magical,mathy2012s,gobet2001chunking,simon1974big}. Miller’s work~\cite{miller1956magical} and subsequent studies suggest that humans can reliably process on the order of $7 \pm 2$ discrete information units, often referred to as ``chunks''~\cite{miller1956magical,mathy2012s}. We use this range as a reference when selecting values of $n_{\text{bins}}$ for numerical experiments.

\section*{Results}
In this section, we present results from Monte Carlo simulations performed across a range of parameter values. 
We begin by examining the effect of social group size ($l$) on emergent population-level opinion patterns, followed by an analysis of how cognitive compression (as controlled by $n_{\text{bins}}$) influences the stability and structure of opinion clusters.

\begin{figure}[h]
    \centering
    \includegraphics[width=\textwidth]{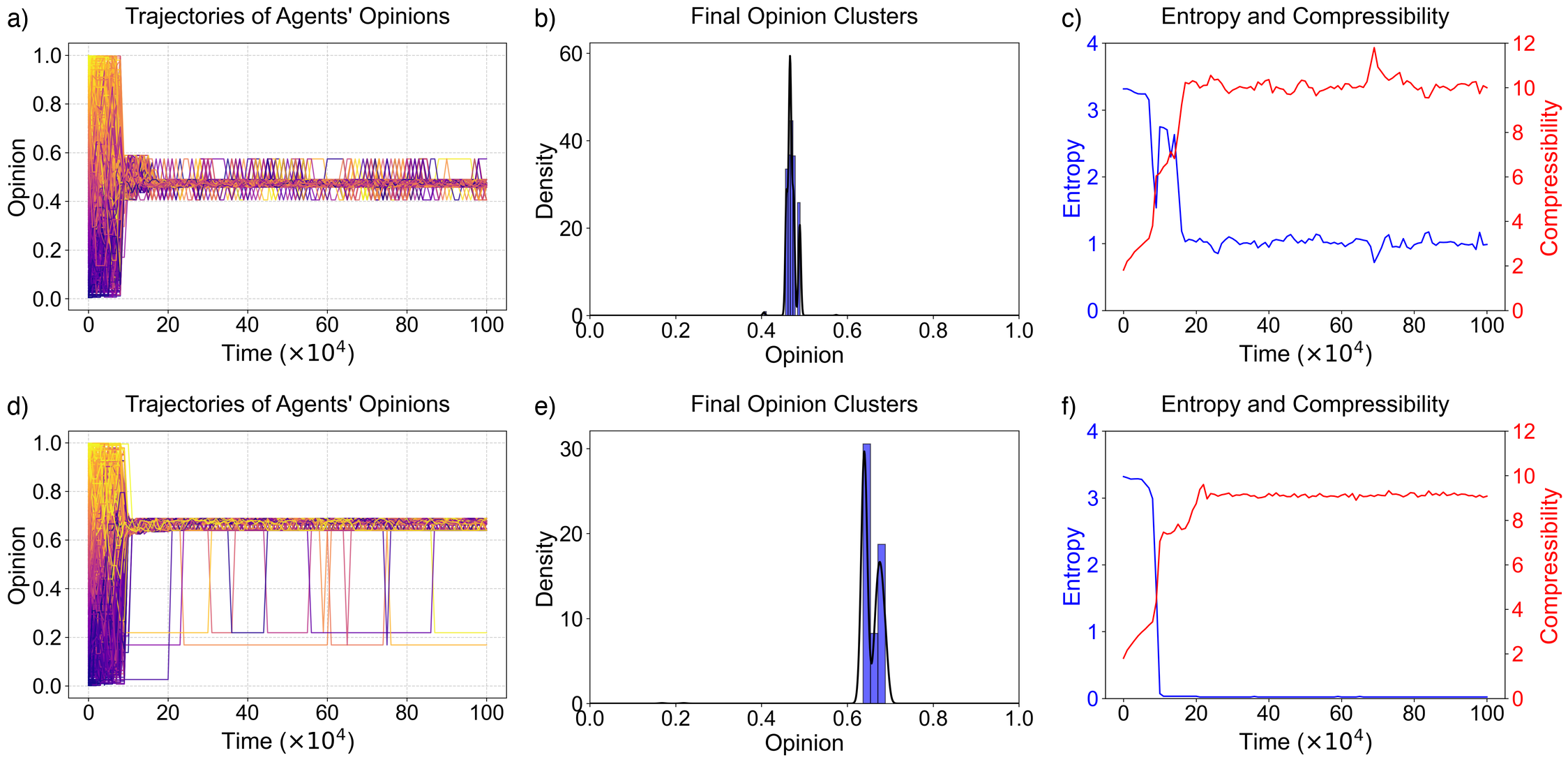}
    \caption{Two realizations of the model, both initialized with the same parameter values but different initial conditions. The top plots correspond to the first realization, while the bottom plots correspond to the second realization. \textbf{(a, d)} Evolution of individual opinions over time, showing the trajectories of each agent's opinion for runs 1 and 2, respectively. The trajectories for each node are colored according to their initial value.  \textbf{(b, e)} Final opinion clusters for runs 1 and 2. \textbf{(c, f)} Evolution of entropy (blue) and compressibility (red) of the entire system. Simulation parameters: $N_{\text{nodes}} = 1000$, $n_{\text{bins}} = 10$ and local group size $= 300$.}
    \label{fig:typical_outcome}
\end{figure}

A typical outcome of the simulation is shown in Fig~\ref{fig:typical_outcome}. In Fig~\ref{fig:typical_outcome}(a) the evolution of agents’ opinions over the course of the simulation is shown. There each line corresponds to the trajectory of the opinion of individual agent and where the color of the line is assigned based on agent's initial opinion at the start of the simulation. This allows us to see how far each agent’s opinion travels in the opinion space. Trajectories of agents with initial opinions close to zero are shown in purple shades, while those starting near one are in yellow shades. With these model parameters, we observe that agents’ opinions shift considerably before eventually forming a single, well-defined cluster, while a small number of agents remain outside the main group. A notable feature of the model is the variability of opinions within clusters, which distinguishes it from previous models. Agents do not converge on the exact same opinion within a group; instead, we observe a distribution of opinions within each cluster, as shown in Fig~\ref{fig:typical_outcome}(b). Moreover, agents continue to adjust their opinions while remaining within the cluster. Occasionally, a few agents shift from the main cluster to outlier positions, and vice versa. In Fig~\ref{fig:typical_outcome}(c), we plot the evolution of entropy and compressibility measures for the global opinion distribution. These reveal that entropy decreases as opinion clusters form and the overall distribution becomes more compressible.

Fig~\ref{fig:typical_outcome}(d,e,f) show the same metrics for another simulation of the model with the same parameter values but a different initial opinion distribution (both drawn from a uniform distribution). Once again, we observe the formation of a single main opinion cluster with internal variability and a few outlier opinions that occasionally move in and out of the main group. However, the position of the main cluster and the outliers differs. This illustrates the model’s sensitivity to the initial opinion distribution. 
It is possible that as the number of agents increases, the final position of the opinion clusters will become more robust to the choice of initial opinion distribution (in the examples discussed, the total number of agents in the system is 1000).

{                                                                                                   
  Since we start with uniformly distributed opinions drawn from the continuous distribution, roughly speaking we
  had around 1000 unique opinions in the beginning of the simulation. As the model evolves, due to swapping       
  mechanisms, the number of unique opinions decreases (or remains constant), with only a small number of unique   
  opinions surviving by the time a stable cluster is formed.                                                      
  }                                                               

  \begin{figure}[ht]
        \centering
        \includegraphics[width=0.6\textwidth]{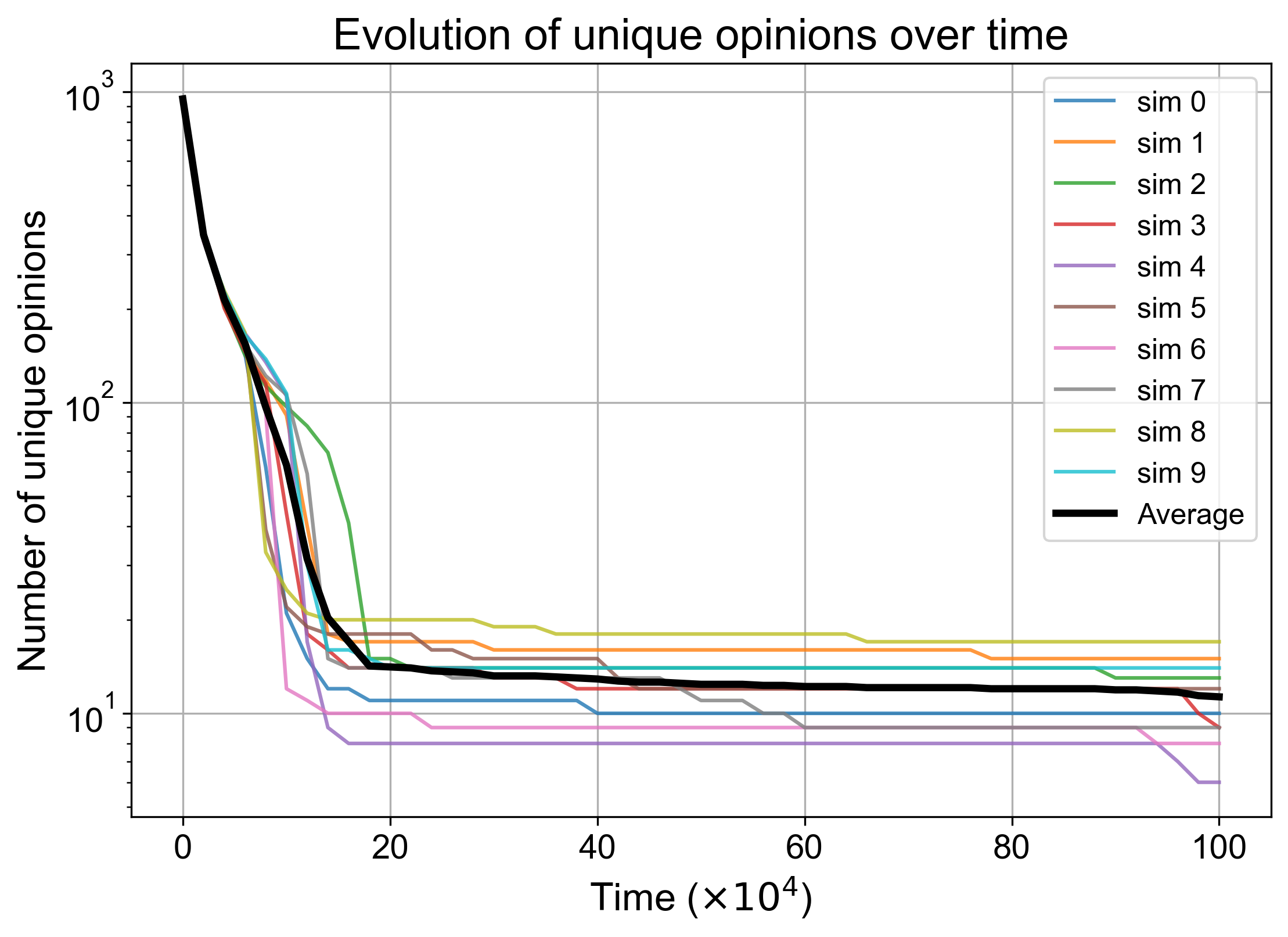}
        \caption{Evolution of the number of unique opinions over time for each of the 10 independent simulation runs (coloured lines), together with their average (black line). Simulation parameters: $N = 1000$ agents, $n_{\mathrm{bins}} = 10$, local group size $l = 300$.}
        \label{fig:unique_opinions_per_sim}
    \end{figure}

  {
    To quantify how opinion diversity evolves over the course of the simulation,
    we track the number of unique opinion present in the population over time
    (Fig~\ref{fig:unique_opinions_per_sim}). We performed 10 independent runs, each starting from opinions drawn uniformly at random from $[0, 1]$ and evolving for $10^6$ steps. All runs exhibit a
   consistent decay in the number of surviving unique opinions and converge toward a small number (of the order of 10). The logarithmic scale on the $y$-axis illustrates an approximately exponential decline at early times, which slows as the opinion clusters form.
  }

\subsection*{Effect of social group size on the macro-level opinion distribution}\label{sec:local_group_size_effect}

In Fig~\ref{fig:typical_outcome}, the local group size was set to 300, which is considered large in our model. For such a large social group, the model tends to settle into what can be described as consensus. Next, we examine examples of outcomes produced by smaller local group sizes.

\begin{figure}[h]
    \centering
    \includegraphics[width=\textwidth]{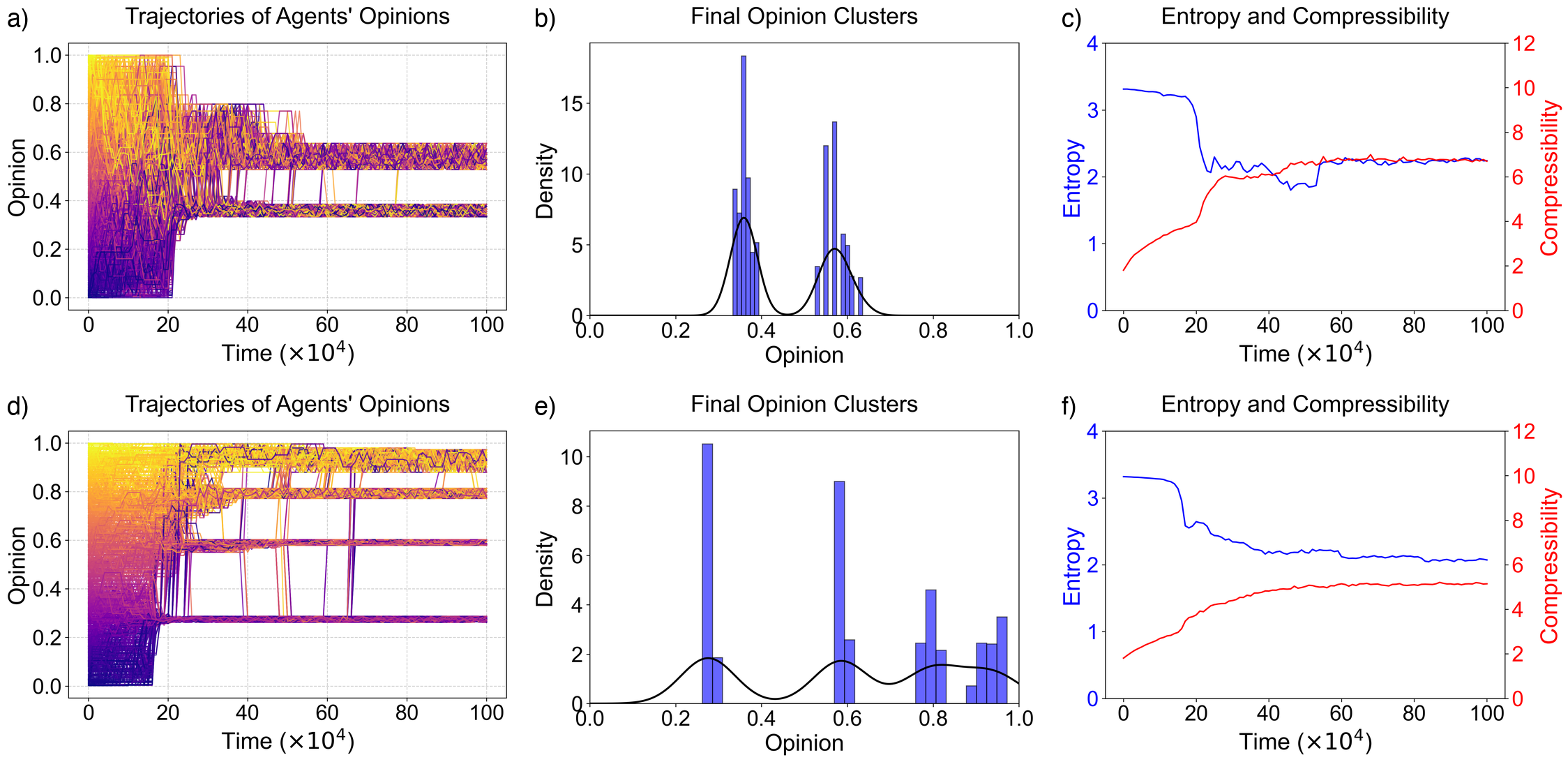}
    \caption{Typical outcome for local group size ($l$) equal to 200 \textbf{(a,b,c)} and 100 \textbf{(d,e,f)}. \textbf{(a, d)} Evolution of individual opinions over time, showing the trajectories of each agent's opinion for $l=200$ and $l=100$, respectively. The trajectories for each node are colored according to their initial value.  \textbf{(b, e)} Final opinion clusters for $l=200$ and $l=100$. \textbf{(c, f)} Evolution of entropy (blue) and compressibility (orange) of the entire system. Simulation parameters: $N_{\text{nodes}} = 1000$, $n_{\text{bins}} = 10$.}
    \label{fig:l_200_100}
\end{figure}

Fig~\ref{fig:l_200_100} shows typical model outcomes for local group sizes ($l$) of 200 and 100. For $l = 200$, we observe the formation of two well-defined clusters, which illustrates opinion polarization. Agents continue to move within opinion clusters, with occasional jumps between them, and variability persists within each group. As the group size decreases further ($l = 100$), more opinion clusters emerge---an outcome we describe as opinion fragmentation.

\blue{Now that we have seen that the in single realizations of the model produces clusters; we want to assess how the emergence of the number of clusters changes as we vary group size parameter, $l$. Naturally, as these simulations are stochastic in nature we are not certain to get the exact same behavior, even with the exact same initial conditions. This variability means that we may observe a different number of clusters emerging for the same parameter set. However, we are interested in the general trend over multiple realizations of the simulation.}

\begin{figure}
    \centering
    \includegraphics[width=0.9\linewidth]{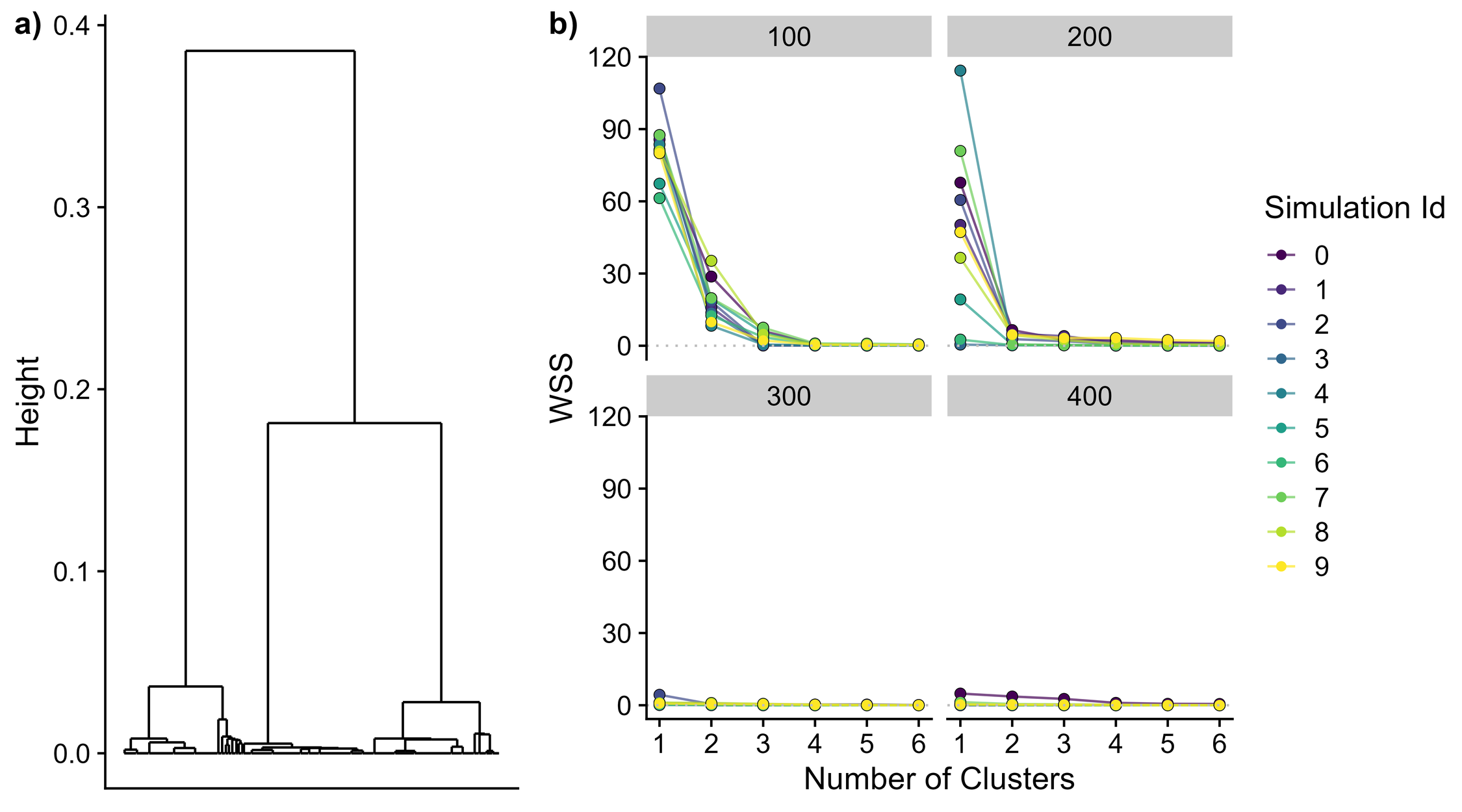}
    \caption{a) An example dendrogram, in which we can agents are group together via their proxomity in opinion space. In a dendrogram we can recover the cluster membership for the agents by ``cutting" the tree a given height. b) Elbow plots of the within-sum of squares found when the number of clusters is fixed across different simulation runs using hierarchical clustering with single linkage. Panels show different group sizes, $l$, of 100, 200, 300 and 400. Simulation parameters: $N_{\text{nodes}} = 1000$ and $n_{\text{bins}} = 10$.}
    \label{fig:varying_l_finding_no_cl}
\end{figure}

\blue{To allow us to find the number of clusters in a semi‑automated way, we use hierarchical clustering with single linkage \cite{barabasi_network_2016}. Hierarchical clustering is an agglomerative clustering algorithm that iteratively groups agents based on their pairwise Euclidean distances. At each step, the two closest groups are merged, where ``closest" is defined by the smallest distance between any pair of agents belonging to different groups. This process continues until all agents form a single cluster. This procedure produces a tree-like structure, refereed to as a dendrogram, representing the possible cluster membership for each agent if we were to selected a desired number of clusters \cite{barabasi_network_2016}. This cluster membership can be recovered by ``cutting" the tree at a specified height, see Fig~\ref{fig:varying_l_finding_no_cl}a) for an example dendrogram. Once we have this tree, we need a way to identify the optimal number of clusters present in it, i.e., which number of clusters best captures the variability in the data a given $l$, for this we use a Elbow plot \cite{ketchen_jr_application_1996}.}

\blue{A Elbow plot shows how the within‑sum of squares (WSS) varies across different possible cluster numbers. In this setting, the within‑sum of squares (WSS) measures the total squared distance between each agent and the centroid of its assigned cluster. As the number of clusters increases, the WSS decreases, since clusters become smaller and more homogeneous. By examining the rate of decrease in WSS as a function of the number of clusters, we can identify an ``elbow" point where additional clusters provide only marginal improvements in fit. This elbow is taken as an estimate of the appropriate number of clusters. We perform this clustering on the last time point for each simulation and construct the Eblow plot from this data (Fig~\ref{fig:varying_l_finding_no_cl}b)).}

\blue{In Fig~\ref{fig:varying_l_finding_no_cl}b), we can see how the number of clusters varies across simulations for each group size, where, in general, the number of detected clusters decreases as $l$ increases. Examining each panel in Fig~\ref{fig:varying_l_finding_no_cl}b), we see that when the group size is set to 100, we can see that between 3 and 4 clusters is reasonable, while when the group size is set to 200, we observe that two clusters explain almost all the variability within the clusters. Once the group size is set to 300 and 400, it appears that a single cluster is the best explanation of the variability between agents. To see the trajectories for each of these simulations with the clustering solutions, see S2 Appendix.}

We observe a similarity between how social group size acts in our model and the role of the confidence bound parameter in the Deffuant model. In both cases, smaller values of the control parameters result in the formation of large number of clusters. In the Deffuant model, an analytical expression can be derived to predict the number of final opinion clusters as a function of the confidence bound~\cite{pineda2009noisy, dubovskaya2023analysis}. However, unlike in the Deffuant model where opinions are drawn toward nearby clusters, in our model, opinions can end up in clusters located on the opposite side of the opinion space from their starting points.

\blue{We also observed some exotic outcomes. In some simulations with $l = 100$, four clusters form midway through the simulation, but one cluster subsequently dissolves into the others due to occasional jumps between clusters. Similar behavior has been observed in the Deffuant model, where such dissolution is attributed to noise~\cite{pineda2011diffusing}. In some simulations with a large group size (e.g., $l = 500$), no well-defined clusters appeared. Instead, agents kept shifting among a few unique opinions distributed across the entire opinion space.}

\subsection*{Effect of information compression on the macro-level opinion distribution}

The second mechanism we explore is the level of compression people apply when processing diverse information. In the model, this is represented by the number of bins ($n_{\text{bins}}$) used to compress opinion data before estimating its complexity via entropy. A smaller number of bins implies a higher level of compression.

\begin{figure}[h]
    \centering
    \includegraphics[width=\textwidth]{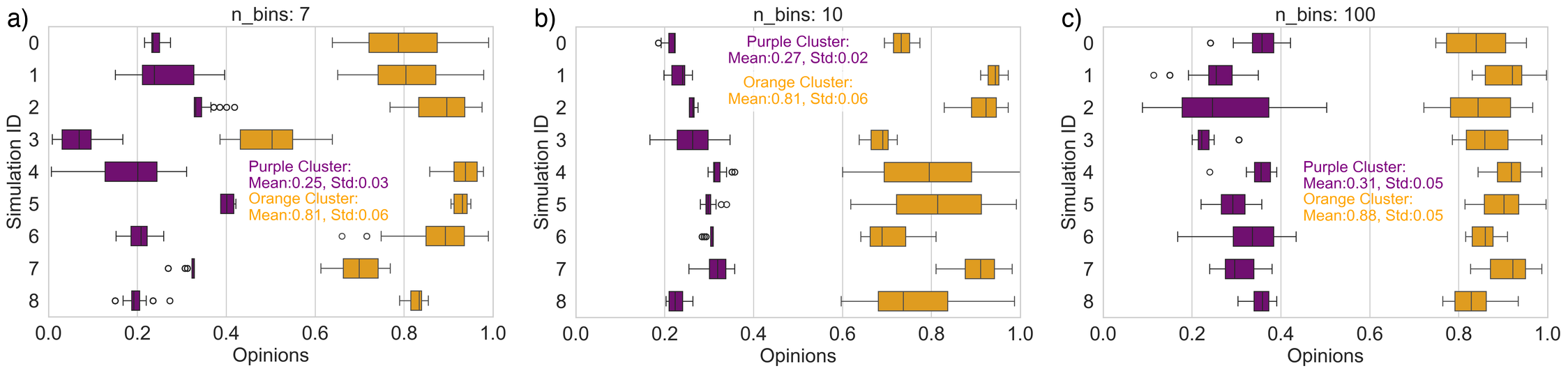}
    \caption{Statistical summary of distribution of opinions for 9 different simulations for different values of parameter $n_{\text{bins}}$ while keeping the other simulation parameters fixed: a) $n_{\text{bins}} = 7$; b) $n_{\text{bins}} = 10$; c) $n_{\text{bins}} = 100$. The boxes correspond to the interquartile range of the cluster, central line represents the median with whiskers extending to the most extreme non-outlier points, and outliers shown as individual points. Each simulation resulted in two distinct clusters, where the first cluster is colored in purple while the second cluster is colored in orange. The average mean and standard deviation (Std) across all purple and orange clusters are also printed. Simulation parameters: $N_{\text{nodes}} = 1000$, and local group size $= 200$.}
    \label{fig:cluster_stats}
\end{figure}

Given the stochastic nature of the model, we investigate how stable the number, structure, and positions of clusters are across simulations as the level of compression changes. To address this, we run a series of simulations with identical parameter values and starting opinion distribution, varying only $n_{\text{bins}}$.

\blue{To analyze the emergent opinion structure, we applied $K$-means clustering~\cite{macqueen1967some} to the final opinion distribution of each simulation run, testing $k = 2$ to $k = 10$. The optimal $k$ was selected by maximizing the silhouette score~\cite{rousseeuw1987silhouettes}}.
% \blue{To enable consistent comparison across simulation runs, clusters were reordered within each trial so that cluster~0 always corresponds to the group with the lowest mean opinion value, cluster~1 to the next lowest, and so on.}
\blue{The resulting cluster assignments are visualized as horizontal box plots (Fig~\ref{fig:cluster_stats}), where each row corresponds to an independent simulation run and each box represents the distribution of opinion values within a single cluster. Boxes are colored by cluster identity (purple for cluster~0, orange for cluster~1). The box plots display the median, interquartile range, and whiskers extending to $1.5 \times \mathrm{IQR}$.}

Fig~\ref{fig:cluster_stats} presents a statistical summary of opinion clusters across nine simulations for each value of $n_{\text{bins}}$. With a local group size of 200, two distinct opinion clusters consistently form. However, when $n_{\text{bins}} = 7$, both the median and the width of the clusters vary noticeably between simulations. As $n_{\text{bins}}$ increases (i.e., compression decreases), both the position and structure of the clusters become more predictable.

Compression level also affects the model’s convergence time. The simulation takes noticeably longer to converge as the number of bins increases and compression decreases, making low-compression scenarios harder to analyze. Nevertheless, we attempted to approximate a no-compression case by using $n_{\text{bins}} = l$ for local entropy estimation and $n_{\text{bins}} = N_{\text{nodes}}$ (where $N_{\text{nodes}}$ is the total number of agents) for global entropy. \blue{We observe two types of outcomes.} In one case, the system remains in a disordered state, similar to the case of very large local group sizes. In the other, we see the formation of equally sized, distinct clusters located on opposite ends of the opinion space. However, due to computational limitations, these results were not thoroughly investigated. 

\section*{Discussion}

In this study, we proposed a novel agent-based model of opinion formation inspired by optimal distinctiveness theory and theory of cognitive compression. The model explores how individuals’ competing desires—to be unique within their social group while maintaining a simplified perception of global opinion distributions—can shape population-level patterns such as opinion group formation and polarization. By incorporating cognitive compression alongside these identity-based mechanisms, we aimed to bridge psychological motivations with complex system dynamics.

Our findings demonstrate that the individuals' desire to having balance between local differentiation and global simplicity of the opinion distributions in the population can lead to the formation of opinion groups and polarization. The model captures several empirically observed features of real-world opinion patterns: (i) the emergence of distinct opinion groups; (ii) variability of opinions within these groups; and (iii) ongoing opinion mobility, even after group structures have formed. These results suggest that investigated individual-level cognitive and social mechanisms can maintain diversity and dynamism within opinion groups, helping to explain why real-world polarization rarely results in total consensus or strictly homogeneous clusters.

With computational experiments performed on the model we found that individual’s social group size plays a critical role in shaping collective outcomes. When social group sizes are moderate (consistent with Dunbar’s number), opinion groups form  leading to consensus, polarization or fragmentation in opinions. These findings provide a mechanistic basis for understanding how the size of individuals' social group constrains or promotes polarization.

We also examined the role of cognitive compression (the degree to which individuals simplify complex information). Higher compression levels increased path dependence in simulations and led to greater unpredictability in the positioning of opinion clusters. Lower compression resulted in more stable and consistent opinion clusters across multiple simulations. These outcomes underscore how cognitive constraints can influence the predictability and variability of population-level opinion distribution.

To isolate the effects of these psychological and cognitive mechanisms, we intentionally adopted a minimal model, including the simplifying assumption of a fully connected population. Since it is well established that social network structure can significantly affect opinion dynamics~\cite{fennell2021generalized, dubovskaya2025modeling, meng2018opinion, karsai2011small}, an important direction for future work is to incorporate heterogeneous network structures into the model. This would allow us to examine how various topologies, e.g., clustered, modular, or small-world networks, affect population-level opinion patterns.

Another promising extension of the model is to give agents memory, so that their perception of the global opinion distribution would be limited by their past interactions and memory capacity. In the current model, individuals have access to the full population-level opinion distribution when making a decision about changing their own opinion. In reality, however, individuals only encounter a subset of opinions shaped by past interactions and cognitive limitations. Incorporating limited memory to the agents could reveal how limited exposure affects opinion group formation and stability.

In conclusion, we introduced a new opinion formation model that integrates cognitive compression with identity-driven motivations for uniqueness and inclusion. This framework offers a plausible psychological basis for the emergence of diverse yet structured opinion clusters in society. The model is both simple enough to yield insight into fundamental mechanisms and general enough to be of interest across disciplines including psychology, sociology, mathematics, and complexity science. We hope this work will encourage further interdisciplinary research into the cognitive and social foundations of polarization, and inspire future models that connect individual motivations with emergent collective behavior in more realistic social settings. 

\section*{Acknowledgments}
This publication has emanated from research conducted with the financial support of Taighde Éireann – Research Ireland under Grant number 13/RC/2094\_2, and by ERC grant (ID-COMPRESSION, grant number: 101124175). Funded in part by the European Union. Views and opinions expressed are however those of the author(s) only and do not necessarily reflect those of the European Union or the European Research Council Executive Agency. Neither the European Union nor the granting authority can be held responsible for them.

\section*{Supporting information}

\paragraph*{S1 Appendix.}
\label{S1_Appendix}
{\bf ODD Description of the Model.}
This appendix describes the agent-based model using the ODD (Overview, Design concepts, Details) protocol~\cite{grimm2020odd}.

\subsection*{Overview}

\subsubsection*{Purpose and patterns}
The model investigates how cognitively motivated information compression shapes the emergence of collective opinion patterns. Agents update their opinions by balancing two drives: increasing opinion diversity within their local social group and reducing complexity in the global opinion distribution. The model is designed to reproduce the spontaneous formation of opinion clusters from an initially uniform distribution.

\subsubsection*{Entities, state variables, and scales}
The model contains $N = 1000$ agents. Each agent $i$ holds a single continuous opinion $x_i \in [0, 1]$ and is associated with a local group $L_i$ consisting of the $l$ agents whose opinions are nearest to $x_i$. The model has two control parameters: the local group size $l$ and the number of bins $n_{\text{bins}}$ used in entropy calculations. Time proceeds in discrete Monte Carlo steps.

\subsubsection*{Process overview and scheduling}
At each time step, two agents $i$ and $j$ are selected uniformly at random. Agent $i$ tentatively adopts the opinion of agent $j$ and evaluates whether this change increases the ratio of local to global Shannon entropy (Eq.~\ref{eq:opinion_update}). If the ratio increases, the opinion change is accepted; otherwise, the original opinion is retained. The simulation runs until a predefined bailout time sufficient for the dynamics to stabilize.

\subsection*{Design concepts}

\subsubsection*{Basic principles}
The model is grounded in two ideas: (1) agents seek to maintain distinctiveness within their local social group (identity differentiation), and (2) agents prefer a globally simpler, more compressible opinion landscape. These drives are operationalized through the ratio of local to global Shannon entropy.

\subsubsection*{Emergence}
Opinion clusters emerge from the repeated application of individual-level update rules. The number, size, and position of clusters are not predetermined but arise from balancing local diversity and global simplicity.

\subsubsection*{Adaptation}
Agents adapt by conditionally switching opinions. The decision criterion compares the local-to-global entropy ratio before and after a hypothetical opinion change.

\subsubsection*{Sensing}
Each agent has access to two pieces of information: the opinion distribution within their local group $L_i$ and the opinion distribution across the entire population. Both distributions are perceived through a discretized (binned) representation, reflecting cognitive chunking~\cite{miller1956magical,gobet2001chunking}.

\subsubsection*{Interaction}
At each step, one agent considers adopting another's opinion. The local group is defined by opinion proximity (homophily), not by an explicit network.

\subsubsection*{Stochasticity}
Initial opinions are drawn from a continuous uniform distribution on $[0, 1]$. Agent pairs are selected uniformly at random at each time step.

\subsubsection*{Observation}
The primary output is the opinion distribution $X(t)$ over time. Key observables include the number, size, and location of opinion clusters, as well as the Shannon entropy of the population-level opinion distribution.

\subsection*{Details}

\subsubsection*{Initialization}
All $N$ agents are assigned opinions independently from $\text{Uniform}(0, 1)$. No pre-existing group structure or network topology is imposed.

\subsubsection*{Input data}
The model uses no external input data.

\subsubsection*{Submodels}

\paragraph{Entropy calculation.}
The opinion space $[0,1]$ is partitioned into $n_{\text{bins}}$ equal-width bins. For a given set of opinions (either the local group $L_i$ or the full population $X$), the proportion of opinions in each bin $k$ is $p_k = n_k / N_{\text{set}}$, where $n_k$ is the count in bin $k$ and $N_{\text{set}}$ is the total number of opinions in the set. The Shannon entropy is $H = -\sum_{k} p_k \log_2 p_k$, with the convention $0 \log_2 0 = 0$. See Algorithm~\ref{alg:entropy}.

\paragraph{Opinion update.}
Given a randomly selected pair $(i, j)$, agent $i$'s local group $L_i$ is identified as the $l$ opinions nearest to $x_i$. The local and global entropies are computed before and after the hypothetical substitution $x_i \leftarrow x_j$. The opinion is updated according to Eq.~\eqref{eq:opinion_update}; see Algorithm~\ref{alg:opinion_update} for pseudocode.
 
\paragraph*{S2 Appendix.}
\label{S2_Appendix}
{\bf Cluster membership for agents given $l$.}
\label{app:cluster_selection}

In the results we showed the results and the general trend for the number of clusters we saw as we varied the group size, $l$. Here we present more information on what each cluster assignment looks like for each simulation. S1--S4 Figs show the cluster membership across all 10 simulation runs for each group size, $l$ of 100, 200, 300 and 400, respectively. It is important for the reader to note that, even with a relatively small number of simulation runs per group size, we observe a consistent change in the number of clusters across the different values of $l$. Due to the computational cost associated with running a larger number of simulations for each group size, we restrict our analysis to these runs; however, the observed changes in cluster structure across group sizes appear to be robust. Here we provide further information on the number of clusters observed within each simulation.

We noted that the optimal number of clusters appeared to be between 3 and 4 for a group size of 100. From S1 Fig, we can see that this is the case, as simulation runs 0, 5, 6 and 8 all appear to be well explained by 4 clusters, while the remainder simulations runs have 3 clusters.

\begin{figure}
\centering
\includegraphics[width=0.99\linewidth]{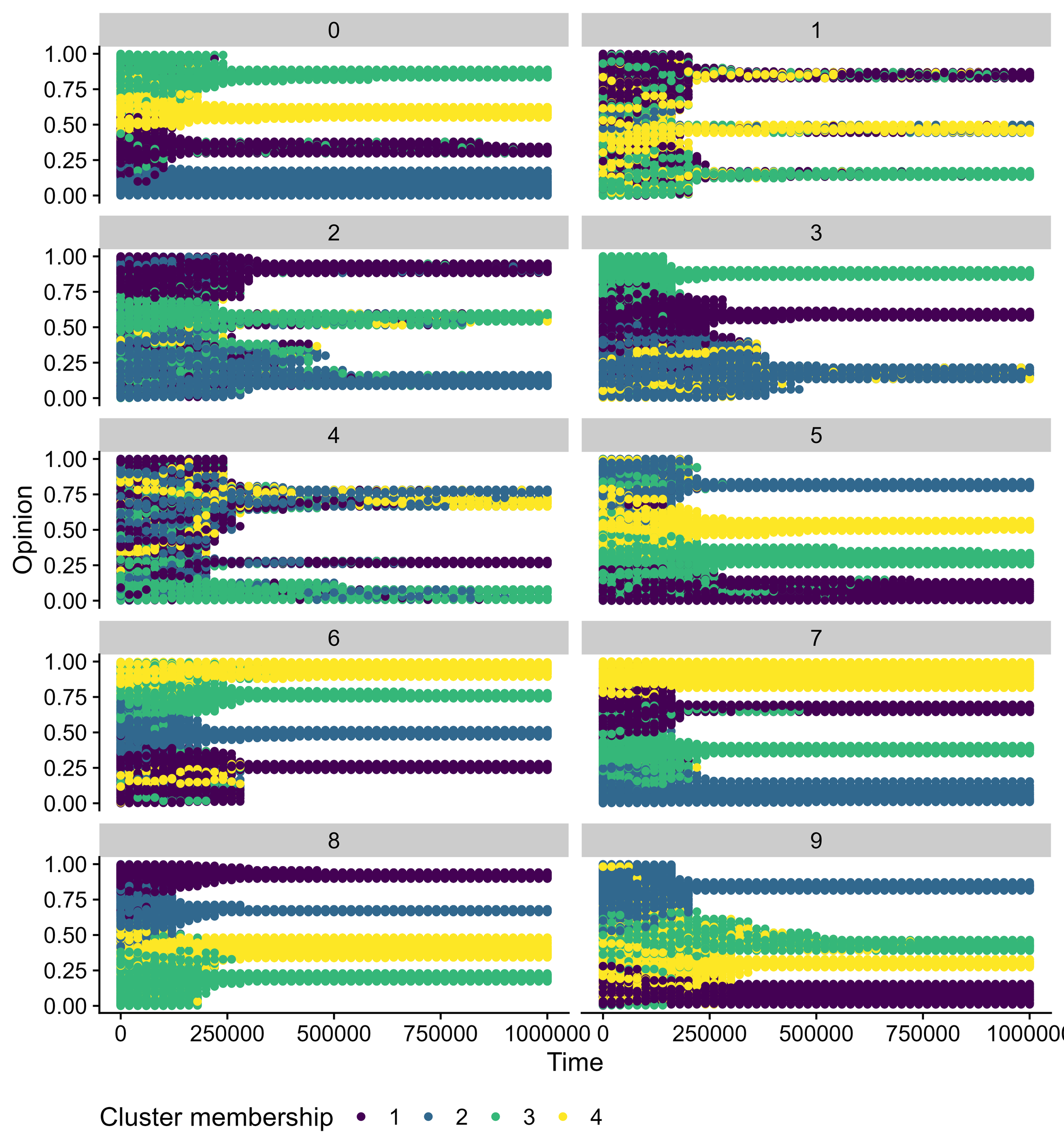}
\caption{{\bf S1 Fig. Cluster membership for each simulation (panels) where $l = 100$.} The color of each agent corresponds to their cluster membership, which was obtained from Sec. \ref{sec:local_group_size_effect}. Here, the number of clusters is selected to be 4.}
\label{fig:app_cls_membership_l=100}
\end{figure}

As we increased the group size, we noted that the optimal number of clusters appeared to be 2 for a group size of 200. From S2 Fig, we can see that this is the case for all simulation runs (with the exception of simulation 2, where only one cluster remains by the end). It is interesting to note that, even though we did find this number of clusters, some of the within-cluster variability does appear to be high. For example, in simulation runs 0 and 9 we observe that cluster 2 has a significantly higher variability between the agents than we see in cluster 1. Ideally, we would study how such clusters would evolve over a longer time horizon, where the cluster variability may reduce, producing more homogeneous clusters, or perhaps not. But due to the computational cost of such experiments, we leave this to future work.

\begin{figure}
\centering
\includegraphics[width=0.99\linewidth]{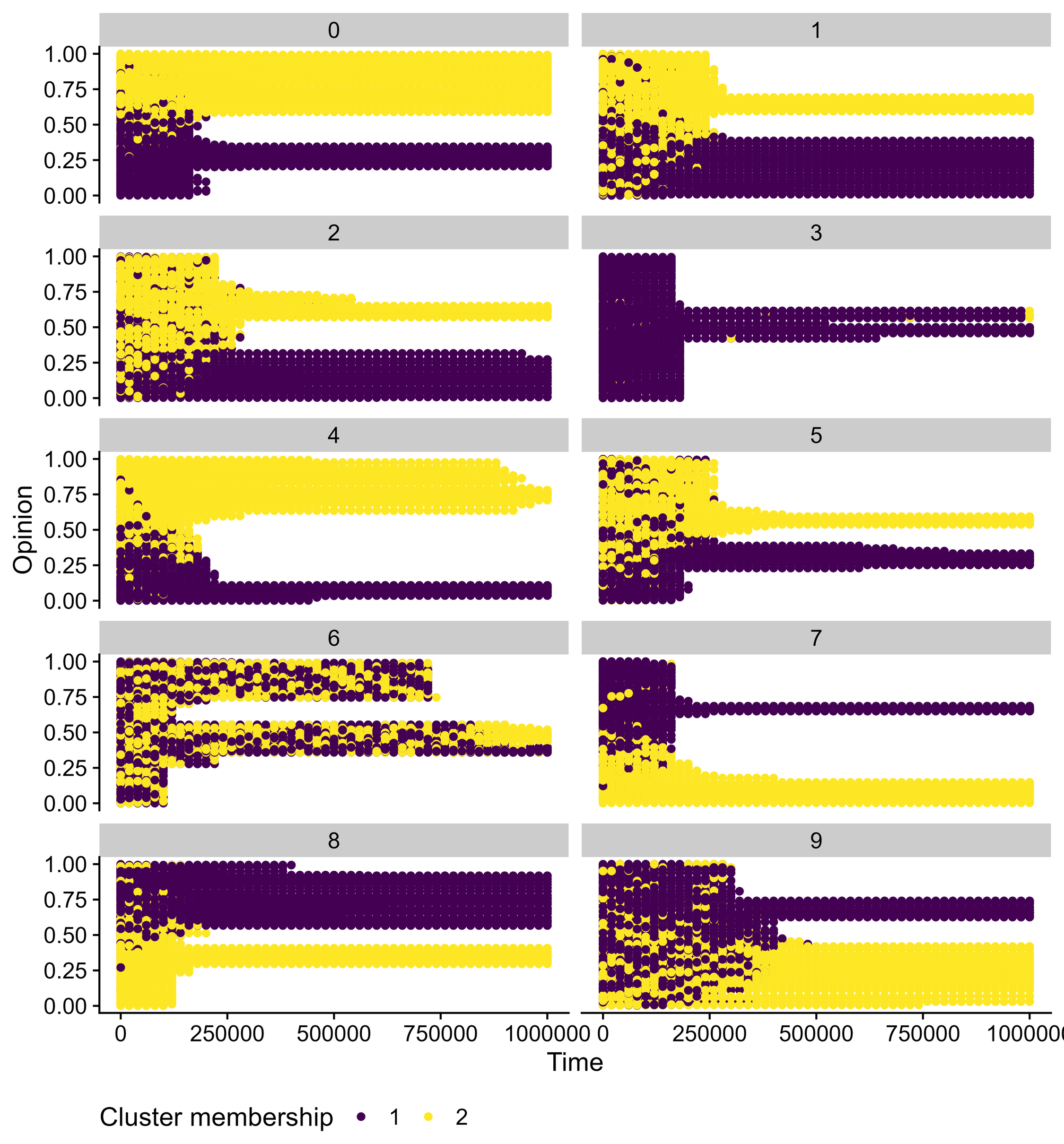}
\caption{{\bf S2 Fig. Cluster membership for each simulation (panels) where $l = 200$.} The color of each agent corresponds to their cluster membership, which was obtained from Sec. \ref{sec:local_group_size_effect}. Here, the number of clusters is selected to be 2.}
\label{fig:app_cls_membership_l=200}
\end{figure}

As we increased the group size further, we noted that the optimal number of clusters appeared to be 1 for group sizes of 300 and 400.
From S3 and S4 Figs, the vast majority of simulations produce a single, well-defined cluster. However, we do observe some interesting behavior in some of the simulation runs, where a small single cluster emerges on a very small set of opinions (see S3 Fig, simulations 0 and 2, and S4 Fig, simulations 1 and 4). Given the group size, one might have expected that these small clusters would merge and not be isolated on these separate distinct, but close, opinions. Again, this warrants further study, but due to computational cost, we leave this to future work.

\begin{figure}
\centering
\includegraphics[width=0.99\linewidth]{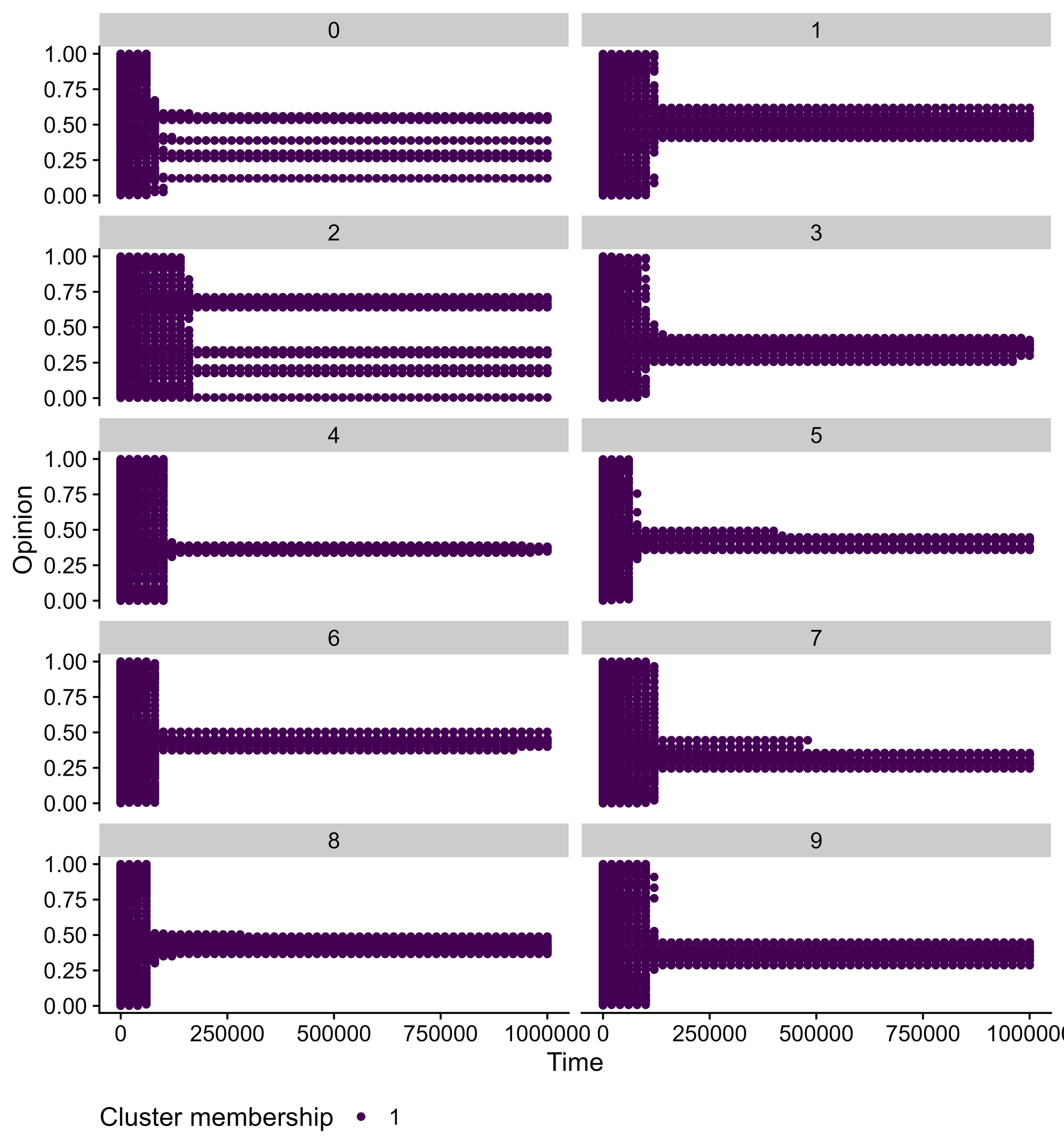}
\caption{{\bf S3 Fig. Cluster membership for each simulation (panels) where $l = 300$.} The color of each agent corresponds to their cluster membership, which was obtained from Sec. \ref{sec:local_group_size_effect}. Here, the number of clusters is selected to be 1.}
\label{fig:app_cls_membership_l=300}
\end{figure}

\begin{figure}
\centering
\includegraphics[width=0.99\linewidth]{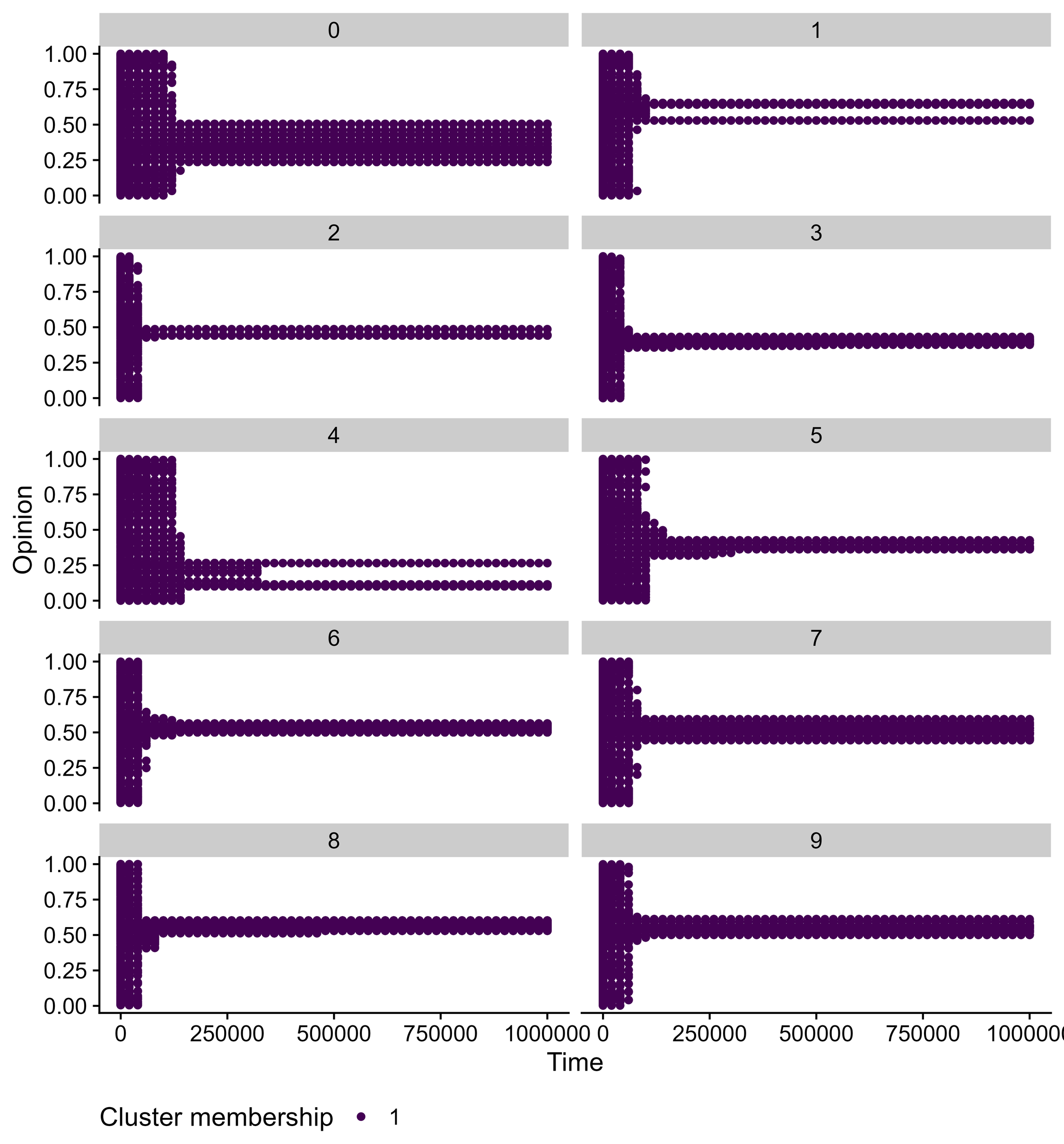}
\caption{{\bf S4 Fig. Cluster membership for each simulation (panels) where $l = 400$.} The color of each agent corresponds to their cluster membership, which was obtained from Sec. \ref{sec:local_group_size_effect}. Here, the number of clusters is selected to be 1.}
\label{fig:app_cls_membership_l=400}
\end{figure} 
%\nolinenumbers

% arXiv: use pre-compiled .bbl instead of running bibtex


\begin{thebibliography}{10}

\bibitem{dellaposta2020pluralistic}
DellaPosta D.
\newblock Pluralistic collapse: The ``oil spill'' model of mass opinion
  polarization.
\newblock American Sociological Review. 2020;85(3):507--536.

\bibitem{Graham2020}
Graham MH, Svolik MW.
\newblock Democracy in America? Partisanship, Polarization, and the Robustness
  of Support for Democracy in the United States.
\newblock American Political Science Review. 2020;114(2):392--409.
\newblock doi:{10.1017/S0003055420000052}.

\bibitem{Arbatli2021}
Arbatli E, Rosenberg D.
\newblock United We Stand, Divided We Rule: How Political Polarization Erodes
  Democracy.
\newblock Democratization. 2021;28(2):285--307.
\newblock doi:{10.1080/13510347.2020.1818068}.

\bibitem{Zhou2016}
Zhou J.
\newblock Boomerangs Versus Javelins: How Polarization Constrains Communication
  on Climate Change.
\newblock Environmental Politics. 2016;25(5):788--811.
\newblock doi:{10.1080/09644016.2016.1166602}.

\bibitem{druckman2013elite}
Druckman JN, Peterson E, Slothuus R.
\newblock How elite partisan polarization affects public opinion formation.
\newblock American Political Science Review. 2013;107(1):57--79.
\newblock doi:{10.1017/S0003055412000500}.

\bibitem{mason2018ideologues}
Mason L.
\newblock Ideologues without issues: The polarizing consequences of ideological
  identities.
\newblock Public Opinion Quarterly. 2018;82(S1):866--887.
\newblock doi:{10.1093/poq/nfy005}.

\bibitem{kubin2021role}
Kubin E, Von~Sikorski C.
\newblock The role of (social) media in political polarization: a systematic
  review.
\newblock Annals of the International Communication Association.
  2021;45(3):188--206.

\bibitem{mutz2018status}
Mutz DC.
\newblock Status threat, not economic hardship, explains the 2016 presidential
  vote.
\newblock Proceedings of the National Academy of Sciences.
  2018;115(19):E4330--E4339.
\newblock doi:{10.1073/pnas.1718155115}.

\bibitem{davis2015theories}
Davis R, Campbell R, Hildon Z, Hobbs L, Michie S.
\newblock Theories of behaviour and behaviour change across the social and
  behavioural sciences: a scoping review.
\newblock Health psychology review. 2015;9(3):323--344.

\bibitem{gavrilets2024modelling}
Gavrilets S, Tverskoi D, S{\'a}nchez A.
\newblock Modelling social norms: an integration of the norm-utility approach
  with beliefs dynamics.
\newblock Philosophical Transactions of the Royal Society B.
  2024;379(1897):20230027.

\bibitem{galesic2023beyond}
Galesic M, Barkoczi D, Berdahl AM, Biro D, Carbone G, Giannoccaro I, et~al.
\newblock Beyond collective intelligence: Collective adaptation.
\newblock Journal of the Royal Society interface. 2023;20(200):20220736.

\bibitem{centola2015spontaneous}
Centola D, Baronchelli A.
\newblock The spontaneous emergence of conventions: An experimental study of
  cultural evolution.
\newblock Proceedings of the National Academy of Sciences.
  2015;112(7):1989--1994.

\bibitem{castellano2009statistical}
Castellano C, Fortunato S, Loreto V.
\newblock Statistical physics of social dynamics.
\newblock Reviews of modern physics. 2009;81(2):591--646.

\bibitem{warncke2025country}
Warncke P.
\newblock What Explains Country-Level Differences in Political Belief System
  Coherence?
\newblock Political Behavior. 2025;47:1853--1876.
\newblock doi:{10.1007/s11109-025-10015-9}.

\bibitem{luders2024attitude}
L{\"u}ders A, Carpentras D, Quayle M.
\newblock Attitude networks as intergroup realities: Using network-modelling to
  research attitude-identity relationships in polarized political contexts.
\newblock British Journal of Social Psychology. 2024;63(1):37--51.

\bibitem{o2024strategic}
O’Reilly C, Mannion S, Maher PJ, Smith EM, MacCarron P, Quayle M.
\newblock Strategic attitude expressions as identity performance and identity
  creation in interaction.
\newblock Communications Psychology. 2024;2(1):27.

\bibitem{miranda2024indirect}
Miranda M, Pereda M, S{\'a}nchez A, Estrada E.
\newblock Indirect social influence and diffusion of innovations: An
  experimental approach.
\newblock PNAS nexus. 2024;3(10):pgae409.

\bibitem{maher2020mapping}
Maher PJ, MacCarron P, Quayle M.
\newblock Mapping public health responses with attitude networks: the emergence
  of opinion-based groups in the UK’s early COVID-19 response phase.
\newblock British Journal of Social Psychology. 2020;59(3):641--652.

\bibitem{carpentras2022mapping}
Carpentras D, L{\"u}ders A, Quayle M.
\newblock Mapping the global opinion space to explain anti-vaccine attraction.
\newblock Scientific reports. 2022;12(1):6188.

\bibitem{dinkelberg2021multidimensional}
Dinkelberg A, O'Reilly C, MacCarron P, Maher PJ, Quayle M.
\newblock Multidimensional polarization dynamics in US election data in the
  long term (2012--2020) and in the 2020 election cycle.
\newblock Analyses of Social Issues and Public Policy. 2021;21(1):284--311.

\bibitem{chen2025broken}
Chen Y, Speer A, de~Bruin B, Carpentras D, Warncke P.
\newblock A ``broken egg'' of {U.S.} Political Beliefs: Using response-item
  networks ({ResIN}) to measure ideological polarization.
\newblock Network Science. 2025;13:e20.
\newblock doi:{10.1017/nws.2025.10016}.

\bibitem{karsai2016local}
Karsai M, I{\~n}iguez G, Kikas R, Kaski K, Kert{\'e}sz J.
\newblock Local cascades induced global contagion: How heterogeneous
  thresholds, exogenous effects, and unconcerned behaviour govern online
  adoption spreading.
\newblock Scientific reports. 2016;6(1):27178.

\bibitem{pena2025finding}
Pena CB, MacCarron P, O’Sullivan DJ.
\newblock Finding polarized communities and tracking information diffusion on
  Twitter: a network approach on the Irish Abortion Referendum.
\newblock Royal Society Open Science. 2025;12(1):240454.

\bibitem{deffuant2000mixing}
Deffuant G, Neau D, Amblard F, Weisbuch G.
\newblock Mixing beliefs among interacting agents.
\newblock Advances in Complex Systems. 2000;3(01n04):87--98.

\bibitem{degroot1974reaching}
DeGroot MH.
\newblock Reaching a consensus.
\newblock Journal of the American Statistical association.
  1974;69(345):118--121.

\bibitem{axelrod1997dissemination}
Axelrod R.
\newblock The dissemination of culture: A model with local convergence and
  global polarization.
\newblock Journal of conflict resolution. 1997;41(2):203--226.

\bibitem{galesic2021integrating}
Galesic M, Olsson H, Dalege J, Van Der~Does T, Stein DL.
\newblock Integrating social and cognitive aspects of belief dynamics: towards
  a unifying framework.
\newblock Journal of the Royal Society Interface. 2021;18(176):20200857.

\bibitem{friedkin2016network}
Friedkin NE, Proskurnikov AV, Tempo R, Parsegov SE.
\newblock Network science on belief system dynamics under logic constraints.
\newblock Science. 2016;354(6310):321--326.

\bibitem{parsegov2015new}
Parsegov SE, Proskurnikov AV, Tempo R, Friedkin NE.
\newblock A new model of opinion dynamics for social actors with multiple
  interdependent attitudes and prejudices.
\newblock In: 2015 54th IEEE Conference on Decision and Control (CDC). IEEE;
  2015. p. 3475--3480.

\bibitem{flache2017models}
Flache A, M{\"a}s M, Feliciani T, Chattoe-Brown E, Deffuant G, Huet S, et~al.
\newblock Models of social influence: Towards the next frontiers.
\newblock Jasss-The journal of artificial societies and social simulation.
  2017;20(4):2.

\bibitem{vallacher2017computational}
Vallacher RR, Read SJ, Nowak A.
\newblock Computational social psychology.
\newblock Routledge; 2017.

\bibitem{hegselmann2005opinion}
Hegselmann R, Krause U.
\newblock Opinion dynamics driven by various ways of averaging.
\newblock Computational Economics. 2005;25:381--405.

\bibitem{pineda2009noisy}
Pineda M, Toral R, Hernandez-Garcia E.
\newblock Noisy continuous-opinion dynamics.
\newblock Journal of Statistical Mechanics: Theory and Experiment.
  2009;2009(08):P08001.

\bibitem{pineda2011diffusing}
Pineda M, Toral R, Hern{\'a}ndez-Garc{\'\i}a E.
\newblock Diffusing opinions in bounded confidence processes.
\newblock The European Physical Journal D. 2011;62:109--117.

\bibitem{bliuc2021online}
Bliuc AM, Bouguettaya A, Felise KD.
\newblock Online intergroup polarization across political fault lines: An
  integrative review.
\newblock Frontiers in Psychology. 2021;12:641215.

\bibitem{Quayle10062025}
Quayle M.
\newblock Social identity networks: People holding attitudes are a collective
  social identity information system and bipartite networks are a useful way to
  represent them.
\newblock European Review of Social Psychology. 2025;0(0):1--66.
\newblock doi:{10.1080/10463283.2025.2514433}.

\bibitem{Durrheim06052025}
Durrheim K, Quayle M.
\newblock Human murmuration: Group polarisation as compression in
  interaction-language dynamics captured by large language models.
\newblock European Review of Social Psychology. 2025;0(0):1--40.
\newblock doi:{10.1080/10463283.2025.2499332}.

\bibitem{smith2024polarization}
Smith LG, Thomas EF, Bliuc AM, McGarty C.
\newblock Polarization is the psychological foundation of collective
  engagement.
\newblock Communications psychology. 2024;2(1):41.

\bibitem{snyder2012uniqueness}
Snyder CR, Fromkin HL.
\newblock Uniqueness: The human pursuit of difference.
\newblock Springer Science \& Business Media; 2012.

\bibitem{brewer1991social}
Brewer MB.
\newblock The social self: On being the same and different at the same time.
\newblock Personality and social psychology bulletin. 1991;17(5):475--482.

\bibitem{codol1975so}
Codol JP.
\newblock On the so-called `superior conformity of the self'behavior: Twenty
  experimental investigations.
\newblock European journal of social psychology. 1975;5(4):457--501.

\bibitem{lemaine1974social}
Lemaine G.
\newblock Social differentiation and social originality.
\newblock European Journal of Social Psychology. 1974;4(1):17--52.

\bibitem{pickett2006using}
Pickett CL, Geoffrey J.
\newblock Using Collective Identities for.
\newblock Individuality and the group: Advances in social identity. 2006;
  p.~56.

\bibitem{postmes2006individuality}
Postmes T, Jetten J.
\newblock Individuality and the group: Advances in social identity.
\newblock Sage; 2006.

\bibitem{rubin2012}
Rubin M, Badea C.
\newblock They are all the same!... but for several different reasons: A review
  of the multicausal nature of perceived group variability.
\newblock Current Directions in Psychological Science. 2012;21(6):367--372.

\bibitem{leonardelli2010optimal}
Leonardelli GJ, Pickett CL, Brewer MB.
\newblock Optimal distinctiveness theory: A framework for social identity,
  social cognition, and intergroup relations.
\newblock In: Advances in experimental social psychology. vol.~43. Elsevier;
  2010. p. 63--113.

\bibitem{lippmann1965public}
Lippmann W.
\newblock Public opinion. 1922.
\newblock URL: http://infomotions com/etexts/gutenberg/dirs/etext04/pbp nn10
  htm. 1965;.

\bibitem{tamariz2015culture}
Tamariz M, Kirby S.
\newblock Culture: copying, compression, and conventionality.
\newblock Cognitive science. 2015;39(1):171--183.

\bibitem{chaitin2006limits}
Chaitin G.
\newblock The limits of reason.
\newblock Scientific American. 2006;294(3):74--81.

\bibitem{shannon1948mathematical}
Shannon CE.
\newblock A mathematical theory of communication.
\newblock The Bell system technical journal. 1948;27(3):379--423.

\bibitem{gobet2001chunking}
Gobet F, Lane PC, Croker S, Cheng PC, Jones G, Oliver I, et~al.
\newblock Chunking mechanisms in human learning.
\newblock Trends in cognitive sciences. 2001;5(6):236--243.

\bibitem{miller1956magical}
Miller GA.
\newblock The magical number seven, plus or minus two: Some limits on our
  capacity for processing information.
\newblock Psychological review. 1956;63(2):81.

\bibitem{mathy2018simple}
Mathy F, Chekaf M, Cowan N.
\newblock Simple and complex working memory tasks allow similar benefits of
  information compression.
\newblock Journal of Cognition. 2018;1(1):31.

\bibitem{dunbar2010bondedness}
Dunbar RI, Shultz S.
\newblock Bondedness and sociality.
\newblock Behaviour. 2010; p. 775--803.

\bibitem{dunbar1998social}
Dunbar RI.
\newblock The social brain hypothesis.
\newblock Evolutionary Anthropology: Issues, News, and Reviews: Issues, News,
  and Reviews. 1998;6(5):178--190.

\bibitem{dunbar1992neocortex}
Dunbar RI.
\newblock Neocortex size as a constraint on group size in primates.
\newblock Journal of human evolution. 1992;22(6):469--493.

\bibitem{shultz2010encephalization}
Shultz S, Dunbar R.
\newblock Encephalization is not a universal macroevolutionary phenomenon in
  mammals but is associated with sociality.
\newblock Proceedings of the National Academy of Sciences.
  2010;107(50):21582--21586.

\bibitem{barton1996neocortex}
Barton RA.
\newblock Neocortex size and behavioural ecology in primates.
\newblock Proceedings of the Royal Society of London Series B: Biological
  Sciences. 1996;263(1367):173--177.

\bibitem{lewis2011ventromedial}
Lewis PA, Rezaie R, Brown R, Roberts N, Dunbar RI.
\newblock Ventromedial prefrontal volume predicts understanding of others and
  social network size.
\newblock Neuroimage. 2011;57(4):1624--1629.

\bibitem{powell2012orbital}
Powell J, Lewis PA, Roberts N, Garcia-Finana M, Dunbar RI.
\newblock Orbital prefrontal cortex volume predicts social network size: an
  imaging study of individual differences in humans.
\newblock Proceedings of the Royal Society B: Biological Sciences.
  2012;279(1736):2157--2162.

\bibitem{kanai2012online}
Kanai R, Bahrami B, Roylance R, Rees G.
\newblock Online social network size is reflected in human brain structure.
\newblock Proceedings of the Royal Society B: Biological Sciences.
  2012;279(1732):1327--1334.

\bibitem{gonccalves2011modeling}
Gon{\c{c}}alves B, Perra N, Vespignani A.
\newblock Modeling users' activity on twitter networks: Validation of dunbar's
  number.
\newblock PloS one. 2011;6(8):e22656.

\bibitem{hernando2010unravelling}
Hernando A, Villuendas D, Vesperinas C, Abad M, Plastino A.
\newblock Unravelling the size distribution of social groups with information
  theory in complex networks.
\newblock The European Physical Journal B. 2010;76:87--97.

\bibitem{Johnson1982}
Johnson GA.
\newblock Organizational Structure and Scalar Stress.
\newblock In: Theory and Explanation in Archaeology; 1982. p. 389--421.

\bibitem{Kuijt2000}
Kuijt I.
\newblock People and space in early agricultural villages: Exploring daily
  lives, community size, and architecture in the Late Pre-Pottery Neolithic.
\newblock Journal of Anthropological Archaeology. 2000;19(1):75--102.
\newblock doi:{10.1006/jaar.1999.0342}.

\bibitem{bonabeau2002agent}
Bonabeau E.
\newblock Agent-based modeling: Methods and techniques for simulating human
  systems.
\newblock Proceedings of the national academy of sciences.
  2002;99(suppl\_3):7280--7287.

\bibitem{mathy2012s}
Mathy F, Feldman J.
\newblock What’s magic about magic numbers? Chunking and data compression in
  short-term memory.
\newblock Cognition. 2012;122(3):346--362.

\bibitem{simon1974big}
Simon HA.
\newblock How Big Is a Chunk? By combining data from several experiments, a
  basic human memory unit can be identified and measured.
\newblock Science. 1974;183(4124):482--488.

\bibitem{barabasi_network_2016}
Barabási AL, Pósfai M.
\newblock Network science.
\newblock Cambridge, United Kingdom: Cambridge University Press; 2016.
\newblock Available from: \url{https://networksciencebook.com/}.

\bibitem{ketchen_jr_application_1996}
Ketchen~Jr DJ, Shook CL.
\newblock The {Application} of {Cluster} {Analysis} in {Strategic} {Management}
  {Research}: {An} {Analysis} and {Critique}.
\newblock Strategic Management Journal. 1996;17(6):441--458.
\newblock doi:{10.1002/(SICI)1097-0266(199606)17:6<441::AID-SMJ819>3.0.CO;2-G}.

\bibitem{dubovskaya2023analysis}
Dubovskaya A, Fennell SC, Burke K, Gleeson JP, O`Kiely D.
\newblock Analysis of mean-field approximation for Deffuant opinion dynamics on
  networks.
\newblock SIAM Journal on Applied Mathematics. 2023;83(2):436--459.
\newblock doi:{10.1137/22M1499765}.

\bibitem{macqueen1967some}
MacQueen J.
\newblock Some methods for classification and analysis of multivariate
  observations.
\newblock In: Proceedings of the Fifth Berkeley Symposium on Mathematical
  Statistics and Probability. vol.~1; 1967. p. 281--297.

\bibitem{rousseeuw1987silhouettes}
Rousseeuw PJ.
\newblock Silhouettes: a graphical aid to the interpretation and validation of
  cluster analysis.
\newblock Journal of Computational and Applied Mathematics. 1987;20:53--65.

\bibitem{fennell2021generalized}
Fennell SC, Burke K, Quayle M, Gleeson JP.
\newblock Generalized mean-field approximation for the Deffuant opinion
  dynamics model on networks.
\newblock Physical Review E. 2021;103(1):012314.

\bibitem{dubovskaya2025modeling}
Dubovskaya A, Pena CB, O'Sullivan DJ.
\newblock Modeling diffusion in networks with communities: A multitype
  branching process approach.
\newblock Physical Review E. 2025;111(3):034310.

\bibitem{meng2018opinion}
Meng XF, Van~Gorder RA, Porter MA.
\newblock Opinion formation and distribution in a bounded-confidence model on
  various networks.
\newblock Physical Review E. 2018;97(2):022312.

\bibitem{karsai2011small}
Karsai M, Kivel{\"a} M, Pan RK, Kaski K, Kert{\'e}sz J, Barab{\'a}si AL, et~al.
\newblock Small but slow world: How network topology and burstiness slow down
  spreading.
\newblock Physical Review E—Statistical, Nonlinear, and Soft Matter Physics.
  2011;83(2):025102.

\bibitem{grimm2020odd}
Grimm V, Railsback SF, Vincenot CE, Berger U, Gallagher C, DeAngelis DL, et~al.
\newblock The {ODD} protocol for describing agent-based and other simulation
  models: a second update to improve clarity, replication, and structural
  realism.
\newblock Journal of Artificial Societies and Social Simulation. 2020;23(2):7.
\newblock doi:{10.18564/jasss.4259}.

\end{thebibliography}
\end{document}